\documentclass[twocolumn,showpacs,preprintnumbers,amsmath,amssymb,prl]{revtex4}
\usepackage{graphicx}
\usepackage{dcolumn}
\usepackage{bm}
\usepackage{comment}
\begin{document}

\title{Electron localization in self-assembled Si/Ge(111) quantum
dots}

\author{N. P. Stepina}
\author{A. F. Zinovieva}
\email[]{aigul@isp.nsc.ru}
\author{V. A. Zinovyev}
\author{A. S. Deryabin}
\author{A. V. Dvurechenskii}
\affiliation{ Rzhanov Institute of Semiconductor Physics, Siberian
Branch of the Russian Academy of Sciences, prospekt Lavrent'eva 13,
630090 Novosibirsk, Russia}

\author{L. V. Kulik}
\affiliation{ Institute of Chemical Kinetics and Combustion, SB RAS,
630090 Novosibirsk, Russia}

\date{\today}

\begin{abstract}
{Electron localization in the Si/Ge heterosystem with Si  quantum
dots (QDs)  was studied by  transport and electron spin resonance
(ESR) measurements. For Si QD structures grown on Ge(111)
substrates, the ESR signal with g-factor $g=2.0022\pm0.0001$ and ESR
line width $\Delta H\approx1.2$~Oe was observed and attributed to
the electrons localized in QDs. The g-factor value was explained by
taking into account  the energy band modification due to  strain
effects and quantum confinement.  A strong Ge-Si intermixing in QD
structures grown on Ge(001) is assumed to be main reason of
unobserved ESR signal from QDs. The transport behavior confirms the
efficient electron localization in Si QDs.}
\end{abstract}

\pacs{73.21.La, 72.20.Ee, 76.30.-V }

\maketitle

\section {Introduction}
The long intrinsic electron spin coherence times in Si~\cite{Feher,
Chiba} make Si-based heterostructures a natural choice for future
quantum computation. Especially, an extremely long spin lifetime is
expected in zero-dimensional structures, quantum dots (QDs), due to
strong electron confinement in all three dimensions~\cite{Tahan,
Kroutvar}. One of the most promising Si-based systems with QDs is
Ge/Si heterostructures, in which electrons are always  localized in
Si. In Ge/Si system with self-assembled Ge QDs the band offset makes
QDs the potential barriers for electrons and the electron
localization occurs in Si vicinity of Ge QDs due to the
strain~\cite{Dvur}. Strain-induced localization is not so strong and
for its enhancement the multilayered QD structures were proposed and
realized~\cite{Yakimov}.  In inverse system with Si QDs embedded in
Ge matrix the potential wells for electrons are formed  inside QDs
and effective electron localization can be realized even in a
one-layered  Si QD structure. However,  a small amount of Ge in Si
QDs resulting from GeSi intermixing can frustrate all expectations
of researchers~\cite{Feher2}  leading to shorter spin relaxation
times due to a large spin-orbit interaction in Ge.

Different research groups tried to grow Si QDs on
Ge~\cite{Pachinger, Lee, SionGe111,surfGe001} and faced the problem
of  Ge and Si intermixing caused by strong Ge segregation  during Si
growth on Ge(001)~\cite{wulf}. Moreover, in a Si-grown-on-Ge layer,
dislocations are introduced much earlier~\cite{Pachinger, SionGe111}
than in a Ge-on-Si case, forced to decrease the thickness of
deposited layer below the critical one for island nucleation. Our
previous work~\cite{ste4} demonstrates a fundamental difference in
the QD growth observed for Ge(111) and Ge(001) substrates. Si growth
on Ge(001) is accompanied by strong Ge/Si intermixing even at
sufficiently low ($400-480^\circ$C) temperatures that leads to
formation of two-dimensional (2D) Ge$_x$Si$_{1-x}$ layer. The
transition to three-dimensional (3D) growth is observed after
deposition of  7 Si monolayers (MLs); 9-10 MLs are necessary to
obtain the QD array with density of about $10^{11}$ cm$^{-2}$.  The
average Si content in these layers is $\approx30\%$.  On Ge(111) the
Ge/Si intermixing is strongly suppressed due to a smaller Ge
segregation during Si deposition. As a result, 3D growth of QDs with
high Si content ($\approx88\%$)~\cite{F} is observed without the
underlying 2D layer formation (Volmer-Weber growth). Such high
average Si content in QDs grown on Ge(111) suggests that this type
of QDs can be considered as promising basic elements for future
spintronic devices, provided that electrons are strongly localized
in QDs.

This work is devoted to a study of electron localization in Si QDs
grown on Ge(111)  by transport and electron spin resonance (ESR)
measurements. Conductance measurements demonstrate that transport is
dominated by hopping via localized QD states.  The ESR study reveals
the ESR signal that can be attributed to QD electrons.

The paper begins with a description of experimental structures  in
section II. The theoretical consideration of electron states in QDs,
including the estimation of spin relaxation time and  electron
g-factor, is presented in section III. Section IV demonstrates
electron localization by means the transport measurements.  The
results of ESR study are given in section V.

\section {Experimental details}
Samples were grown  by molecular beam epitaxy on Ge(111) substrates.
Si QDs are formed by 2.5 Si bilayers (BLs) deposition at
$400^\circ$C.  The details of QD formation are described
elsewhere~\cite{ste4}.  The density of QD array obtained by STM is
about $\sim10^{11} $~cm$^{-2}$, the islands have a pyramidal shape
with a hexagonal base, the average lateral size $l$ is about of 15
nm and aspect ratio $h/l\sim0.1$. Raman measurements give the Si
content $\approx88\%$. Cross-sectional transmission electron
microscopy (TEM) displays the dislocation-free structures.

For the transport measurements, the single-layered QD structures
were grown on the Sb-doped Ge buffer layers with variation of Sb
concentration in the range of $\sim 1-5\times 10^{16}$ cm$^{-3}$.
The thickness of Ge cap layer is 40 nm. The Al metal source and
drain electrodes were deposited on the top of the structure and
heated at 350$^\circ$C in the Ar atmosphere to form reproducible
Ohmic contacts. Since the ionization energy of Sb impurities in Ge
is approximately 10 meV and the energies of the electron levels in
Si QDs of this size are sufficiently larger,  at low temperatures
electrons should leave the impurities and fill the levels in QDs.
The resistance along  QD layer was measured by the four-terminal
method. The temperature stability was controlled using a Ge
thermometer.

To increase the response from the sample in ESR experiments, the
structures with five Si QD layers separated by 10 nm Ge spacers were
grown. The structures were capped  with a 0.1~$\mu$m epitaxial n-Si
layer (Sb concentration $\sim5\times10^{16}$cm$^{-3}$). The similar
structures were grown on Ge(001) to verify the possibility of
receiving ESR signal from Si QDs formed on the (001) substrates. The
optimal conditions for QD formation on this type substrates were
used, QDs were grown at $400^\circ$~C by deposition of 10~MLs of Si.
The ESR measurements were performed with a Bruker Elexsys 580 X-band
ESR spectrometer using a dielectric Bruker ER-4118 X-MD-5 cavity.
The sample was glued on a quartz holder, then the entire cavity with
the sample was maintained at a low temperature in a helium flow
cryostat (Oxford CF935).

\section{Electron spin states in a Si QD }

The effective mass  calculations of the energy spectrum, including
strain effects for the electron localized in a Si QD grown on
Ge(111), were performed using the NEXTNANO program~\cite{Next}. Based
on the STM study~\cite{ste4}, the Si truncated cone with base length
$l$=15 nm and height $h$=1.5 nm was used as a Si QD model. The  Si
content was chosen to be 88$\%$ according to the Raman data. For
comparison the energy levels for electron localized in a pure Si QD
were calculated.  The results of calculations show that  ground
state energy $E_{QD}$ for the electron  localized inside a pure Si
QD is 123 meV, while, for Ge$_{0.12}$Si$_{0.88}$  QD, $E_{QD}$=139
meV (energy is measured from conduction band edge $E_c$ in the Si QD
layer). The energy of the first excited state in the pure Si QD is
146 meV (161 meV in the Ge$_{0.12}$Si$_{0.88}$ QD).

The obtained values of confinement  energy are not so large as it is
expected from the conduction band offset   in GeSi heterostructures
$\Delta U_{SiGe}\approx430$~meV (derived from the data of
Refs~\cite{Qteish},~\cite{Weber}). This is a consequence of strain
effect on the band alignment for (111) orientation. The strain in
the Si layer is close to the uniaxial compression along (111)
direction and, opposite to the (001) case, does not induce the
splitting of $\Delta$-valley. The main strain effect results in the
shift of the conduction band edge and corresponding decrease of the
initial band offset down to $\Delta U_{SiGe}\approx240$~meV.

The results of calculations can be used for the estimation of
phonon-assisted spin relaxation time for an electron in the Si QD
ground state. The spin reversal probability
 can be evaluated using the perturbation theory
(PT). The matrix element of spin-phonon interaction
$\langle\uparrow|\hat{H}_{ph}|\downarrow\rangle$ vanishes in the
first order of PT~\cite{Yafet}. The second order contribution is
inversely proportional to the energy gaps between the ground and
excited states. The calculated energy level spacing $\approx$ 20 meV
is one order larger than that for gated QDs~\cite{QD} allowing to
expect a sufficient increase of the spin relaxation time. The spin
relaxation time calculated in the same way as in our previous
papers~\cite{Zin2,relax2,relax2add}, turns out to be
$\sim50$~seconds at temperature 5 K and external magnetic field
$H=3455$~Oe. Such long spin relaxation time is a consequence of both
small spin-orbit interaction in Si and strong electrons confinement
in Si QDs.

With a knowledge of the energy spectrum and strain in the Si QD, it
is possible to   predict the electron $g$-factor value. For
electrons in silicon, the $g$-factor  can be evaluated through PT by
the following equation~\cite{Roth}:
\begin{eqnarray}
g=2\mathbf{I}+\frac{2}{im}\Sigma_{\mu\nu}\frac{1}{E_{0\mu}E_{0\nu}}
\{\mathbf{h}_{0\mu} \mathbf{p}_{\mu\nu}\times \mathbf{p}_{\nu0} +\nonumber \\
+\mathbf{h}_{\nu0} p_{0\mu}\times p_{\mu\nu} + \mathbf{h}_{\mu\nu}
\mathbf{p}_{0\mu}\times \mathbf{p}_{\nu0}\},
\end{eqnarray}
where $\mathbf{I}$ is the unit dyadic, $E_{0\mu}=E_0-E_{\mu}$ is the
energy gap between electron energy level $E_0$ and energy band
$E_{\mu}$ contributing to g-factor correction, $p_{\mu\nu}$ is the
momentum matrix element, and the spin-orbit interaction has the form
of $\mathbf{\mathcal{H}}_{so}=2\mathbf{s}\cdot\mathbf{h} $.

Confinement and strain effects in the Si QD system  lead mainly to
the energy gaps modification, while the matrix elements of the
momentum and spin-orbit interaction change negligibly. Then, for the
estimation of the $g$-factor correction, one needs to know the
positions of the main energy bands and the energy of the electron
localized in a Si QD.

Electron energy $E_0$ can be written as $E_c+E_{QD}$, where $E_c$ is
the conduction band edge and $E_{QD}$ is the energy of the electron
ground state. Then g-factor  correction acquires the form:
\begin{equation}
\delta g\sim\sum_{\mu\nu}
\frac{C_{\mu\nu}}{(E_c+E_{QD}-E_{\mu})(E_c+E_{QD}-E_{\nu})},
\end{equation}
where $C_{\mu\nu}$ is the combination of matrix elements in curly
brackets in Eq.~1. This allows us to separate the strain-induced
shift of band edges $E_c$ and $E_{\mu,\nu}$ and the
confinement-induced change of electron energy level $E_{QD}$.

The $g$-factor value in Si is mainly determined by the contribution
of $\Delta_5$ valence band and $2p$ core states~\cite{Liu}. For
estimation, the shift of these band states  can be taken roughly
equal to the shift of the average energy of three uppermost valence
bands $E_{av}$ (bands of heavy, light and split-off  holes). The
NEXTNANO program allows the calculation of not only the electron
energy in a Si QD, but the  main energy bands position. The obtained
$E_{c}$ and $E_{av}$ values are presented in  Table 1 for a pure Si
QD,  for Ge$_{0.12}$Si$_{0.88}$  QD and for a bulk Si.  The strain
effect results in the upward shifts of $E_{c}$ and $E_{av}$ in a Si
QD relatively to their original values  in a bulk Si, $\Delta E_{c}$ =
0.2 eV and $\Delta E_{av} = 0.113$ eV, correspondingly. Since the
electron energy  is $E_{QD}=0.123$~eV, then the energy gap
$E_{0\mu,\nu}$ determining the correction to the $g$-factor
increases by $\sim0.2$~eV. Such large change of  $E_{0\mu,\nu}$
allows us to expect a sufficient change of the electron $g$-factor
value in Si QDs.

The smallness of $E_{QD}$, as compared to the distance between the
$E_c$ and the nearest valence band at the $\Delta$-point
($E_c-E_{\Delta_5}=4.27$~eV)~\cite{Dargys}, makes it possible to use
the Taylor series expansion of expression $\delta g$ in  parameter
$E_{QD}$ for estimation of $g$-factor value. As a result, we obtain
linear dependence $\delta g\sim K\cdot E_{QD}$, with the coefficient
$K$ being determined by the magnitude of the energy gaps and matrix
elements of momentum and spin-orbit interaction. Similar dependence
of $g$-factor value on the electron binding energy $E_b$ for As, P
and Sb impurities in Si was plotted in the paper by Young {\em et
al.}~\cite{Young} using the empirical data measured by
Feher~\cite{Feher2}. The extrapolation of this linear dependence
allows the authors~\cite{Young} to predict the $g$-factor value for
electrons on conduction band edge $g_{c}$=1.9995. High-precision ESR
measurements have shown an excellent agreement  between the measured
and expected $g_{c}$ value. We suppose that the empirical linear
dependence observed in the work~\cite{Young} is also the result of
Tailor expansion of $\delta g$, but in the small parameter $E_b$.
Then the slope of this line gives the coefficient $K$ for unstrained
bulk Si. In the strained Si this coefficient is changed  due to the
energy bands shift and,  in the simple case of two band
approximation, it takes the value of $K'\sim
C/(E'_c-E'_{\Delta_5})^3$, where $E'_c$ and $E'_{\Delta_5}$ are
strain-modified energy bands positions. It is possible to estimate
$K'$ using the energy bands modification from Table~1  and
coefficient $K$ from the empirical dependence given in the
paper~\cite{Young}.

\begin{table}
\caption{\label{jlab1}The main energy bands positions  and ground
state energy values of the electron in a Si QD calculated using the
NEXTNANO program in the effective mass approximation taking into
account strain effects. The reference point for electron energy was
chosen to be the same as in the work~\cite{Zunger}}. \footnotesize

\begin{ruledtabular}
\begin{tabular*}{3.3in}{@{\extracolsep{\fill}}|c|c|c|c|}
energy band &   pure Si QD &  Ge$_{0.12}$Si$_{0.88}$ QD & Si bulk \\
\hline
$E_c$ & 2.47 & 2.45 & 2.27  \\

$E_{av}$ & 1.203 & 1.246 & 1.09 \\

$E_{QD}+E_c$ & 2.593 & 2.589 & -

\end{tabular*}
\end{ruledtabular}
\end{table}





\begin{figure}
\includegraphics[width=3.3in]{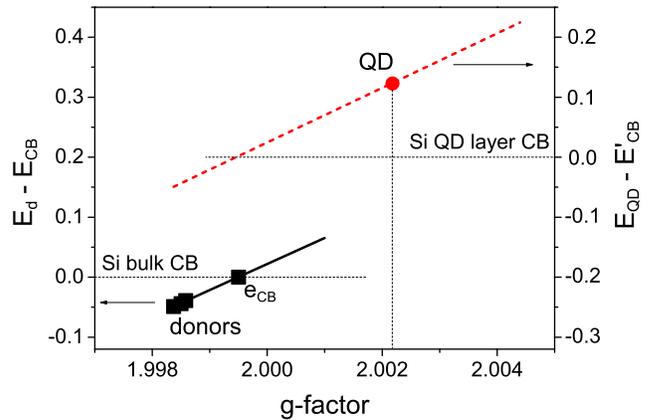}
\caption{\label{Fig.1} An extrapolation of the $g$-factor linear
dependence on the electron binding energy (black solid line)  for
the conduction-band electrons (CB) and three shallow donors  in bulk
Si (filled squares).  The red dashed line is the modification of
this linear dependence for the case of a strained Si QD layer. The
red circle corresponds to the $g$-factor value for a pure Si QD. }
\end{figure}

The black solid line in Fig.~\ref{Fig.1} represents the
extrapolation of the linear dependence of $g$-factor on the electron
binding energy for impurities (filled squares) in bulk unstrained
Si~\cite{Feher2}. The red dash line demonstrates the modification of
this linear dependence for the strained Si QD layer in the simple
case when $g$-factor is determined by the contribution of the only
one nearest energy band. The lines are separated along the energy
axis, because the conduction band edge in the case of the Si QD
layer is shifted to the higher energies due to the strain. The red
dashed line has a slope slightly different from the slope of the
line for the donors in  unstrained bulk Si. The g-factor dependence
is presented for a pure Si QD because the difference between the
slope of the lines plotted for QDs with $100\%$ and $88\%$ Si
content is negligible. The difference will be observed if one takes
into account the change of the constant $C$ due to Ge admixing. For
the electrons localized in a 100$\%$ Si QD, the linear dependence
allows us to predict $g_{QD}=2.00218$ (red circle in
Fig.~\ref{Fig.1}).

\section{Electron localization: transport properties}
Transport measurements in disordered systems allow one to understand
whether the carriers are strongly localized or not. In the case of
strong localization, the  transport is carried out by hopping
between localized states~\cite{Shk84}. In the opposite case of
delocalized carriers, the transport is described by the classical
Drude theory with quantum corrections to the conductivity derived
from the weak localization theory~\cite{gant03}. Different transport
mechanisms manifest themselves in different temperature dependencies
of the conductance and different magnetoresistance behaviors.

\begin{figure}
\includegraphics[width=3.0in]{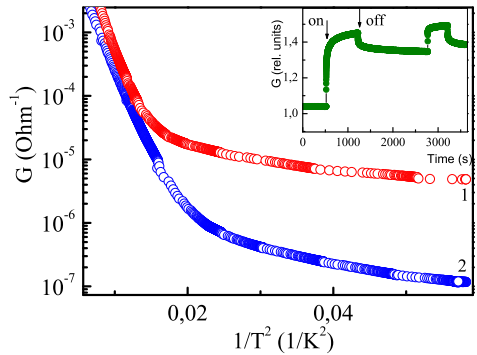}
\caption{\label{Fig.2} Temperature dependencies of the conductance
for QD structures with different doping levels. The photoconductance
kinetics of sample 2 under illumination with the 1.5 $\mu$m
wavelength is shown in the inset.  }
\end{figure}

It was previously shown~\cite{ste9} that the lateral transport in
two-dimensional Ge QD array  with QD sizes and density similar to
the Si QD system under study is described  by variable range hopping
via strong localized QD states  taking into account long-range
Coulomb interaction (Efros-Shklovskii (ES) law~\cite{a22}). To
confirm the strong electron localization in Si QDs and check whether
the mechanism of electron transport in Si QD samples is similar to
the hole transfer in Ge QD samples, the temperature dependence of
conductance, magnetoresistance (MR) and photoconductance were
studied. The temperature dependencies of the lateral conductance for
Si QD samples with two different doping levels are demonstrated in
Fig.~\ref{Fig.2}. The high temperature part of the curves
corresponds to the ionization of impurities to the conduction band.
Analysis of the low temperature parts ($<$10~K) shows that the
temperature dependencies of conductance for both samples are
described  by ES law confirmed the strong carriers localization in
the structure under study. The small temperature of the transition
from the band to hopping transport, as compared to Ge/Si QDs
structures~\cite{ste9}, is explained by smaller ionization energy of
Sb in Ge than in  Si.

An additional heating at $470-600^\circ$C does not change the
conductance of the system. Previously, in the Ge/Si heterostructure
with Ge QDs grown on Si(001), annealing at  $550-625^\circ$C
temperatures resulted in a considerable increase of the conductance
up to the observation of transition from hopping to the diffusion
regime~\cite{ste9}. This was explained by smoothing of the
localization potential and the corresponding increase of the wave
functions overlapping  due to a smearing of the QDs caused by Ge-Si
intermixing. In the structures under study  the diffusive smearing
of Si QDs is suppressed due to a larger stability of Ge/Si
interfaces with (111) orientation than that with
(001)~\cite{Si_Ge111}.

The photoconductance kinetics of sample 2 under illumination with
the 1.5 $\mu$m wavelength light is shown in the inset to
Fig.~\ref{Fig.2}. Slow kinetics under switch on and switch off the
illumination, accompanied with the  persistent photoconductance
effect, was already observed for Ge QD structures and  explained by
the spatial electrons and holes separation in type-II
QDs~\cite{ste5}. We suppose that the same mechanism is responsible
for the observed photoconductance behavior  in the Si QD system.

\begin{figure}
\includegraphics[width=3.0in]{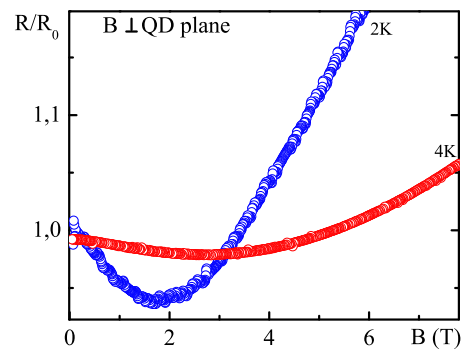}
\caption{\label{Fig.3} Magnetoresistance of Si QD structure in
perpendicular magnetic field at different temperatures. }
\end{figure}

The MR data of Si QD sample 1 measured in the perpendicular magnetic
field at  2~K and 4.2~K are demonstrated in Fig.~\ref{Fig.3}. One
can see  that MR is negative in weak magnetic fields,  then it
crosses  to a positive MR with an increase of the magnetic field.
Recently, we have observed such MR behavior   in the Ge QD
structures grown on Si~\cite{ste7}. The positive MR in high magnetic
fields was explained by suppression of the conductance due to
shrinking of the carrier wave functions, whereas a negative
weak-field MR was referred to the weak localization contributions to
the conductance of clusters with closely located QDs.

Thus, obviously similar transport properties of Ge and Si QD systems
with close structural parameters (QD size, shape and areal density)
allows us to conclude that a strong localization of electrons occurs
in Si QDs as well as a hole localization in Ge QDs.

\section{ESR study }
\begin{figure}
\includegraphics[width=3.3in]{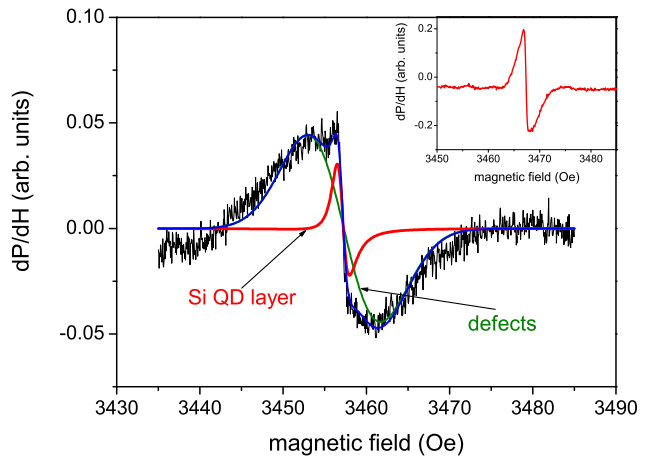}
\caption{\label{Fig.4} The ESR signal obtained for the
heterostructure with 5 layers of Si QDs; microwave power and
frequency P=0.063~mW, $\nu=9.68849$~GHz,  temperature T=4.5K. Solid
lines are the approximation of the experimental signal.  The ESR
signal after annealing, $\nu=9.71679$~GHz, is shown in the inset.}
\end{figure}

The structures with Si QDs grown on Ge(111) were investigated using
ESR spectroscopy at T=4.5 K. The observed  ESR signal  is shown in
Fig.~\ref{Fig.4}. The approximation of the experimental curve
demonstrates that the ESR signal is  a superposition of two ESR
lines - one wide ESR line with the width of $\Delta H\approx6$~Oe
and the second narrow ESR line with $\Delta H\approx1.2$~Oe. Both
signals have isotropic $g$-factor $g=2.0022\pm0.0001$ in spite of
their different origin. The wide ESR line has a Gaussian shape that
indicates a non-homogeneous broadening of the signal. Heating at
650$^\circ$C for 10 min leads to complete disappearance of this ESR
line. A very similar signal was observed in the paper~\cite{stes}
for the oxidized  Ge substrates. We attribute this signal to the
defects in epitaxial Si/Ge layers or in the GeO$_x$ layer. The test
structure grown in the same conditions, but without QDs shows the
only wide ESR signal.

The narrow ESR signal has a Lorentzian shape. Heating does not
affect both ESR line width and g-factor value suggesting that the
signal belongs to the electrons localized in Si QDs. The annealing
stability of the $g$-factor indicates the negligible Ge/Si
intermixing. From the invariance of the ESR line width it can be
assumed that spin relaxation time $T_2$ remains the same (for
Lorentzian lines $\Delta H\sim 1/T_2$). It is well known that the
stronger the carrier confinement, the longer the spin relaxation
time is~\cite{Kroutvar}. It means that the electron localization
radius does not change considerably after annealing that is in a
good agreement with  the results of conductance measurements.

The isotropy of the $g$-factor  does not contradict the Si QD origin
of narrow signal. The $g$-factor anisotropy obtained earlier for QDs
grown on (001) substrates  results from $\Delta$-valley splitting
induced by the uniaxial strain along (001)~\cite{zin1}. In the
structures under study the uniaxial strain along (111) does not
affect the conduction band minima at the $\Delta$-point~\cite{Walle}. The observed isotropic $g$-factor is explained
by missing the $\Delta$-valley splitting for thin Si QD layers grown
on the Ge (111) substrate.

The shape of narrow ESR line is asymmetric and close to the Dysonian
one~\cite{Dyson}  that can be explained by the admixture of
dispersion signal~\cite{Wilam1}. The appearance of noticeable
dispersion contribution is the characteristic feature of the samples
with non-zero conductivity. As shown in our previous work
\cite{Zin3}, the well pronounced asymmetry of ESR line shape   can
be resulted from the hopping transport across the QD array. However,
there is one more possible reason of ESR line asymmetry.
Earlier~\cite{Feher2} for donor electrons in Si the asymmetry of ESR
line was induced by small admixture of Ge atoms ($\sim1\%$). In Si
QD system the asymmetry of ESR line can arise due to penetration of
electron wave function into Ge barrier. It should be noted that  ESR
lines from the defects in Si and Ge usually have symmetric shape and
are characterized by a larger line width, especially for  Ge, in
which ESR line broadening reaches tens of oersteds.

The $g$-factor value $g_{QD}=2.0022$ obtained in ESR experiments is
in a good agreement with  predicted  in Sec.~3 value $g=2.00218$
(red circle in Fig.~\ref{Fig.1}) and is sufficiently larger than
that observed earlier for the electrons in Si. In bulk Si, the
$g$-factor of conduction electrons is $g_{ce}=1.9987$, while for
different quantum-confined systems, $g_{ce}$ varies from 1.9986 to
2.0007~\cite{Graeff, Wilam2}, depending on  the microscopic
structure of quantum wells. A larger value of electron g-factor in
Si QDs  is explained by strong electron confinement  and strong
modification of band structure due to strain.

The fact that the experimental g-factor value  is close to the
predicted one for pure Si QD suggests that QDs in the structures
under study contain pure Si core. This should provide long spin
relaxation times. However, the spin transverse time, determined from
the ESR line width is about of  $T_2\sim 10^{-7}$~s. The
longitudinal spin relaxation time $T_1$ can be estimated from
microwave power dependence of the ESR signal.  In our case the
saturation of the ESR signal does not  observed up to microwave
power $0.4$~mW  that corresponds to the limitation of times
$T_1\leq10 \mu$s. Such short times can be related to spin relaxation
induced by tunneling between QDs. Recently the Dyakonov-Perel
mechanism was proposed for explanation of fast spin relaxation in
dense QD arrays~\cite{zin1}. However, this mechanism should reveal
itself in the anisotropy of the ESR line width that is not observed
in the structure under study. The spin relaxation in Si QD system is
more likely provoked  by intervalley transtions~\cite{Brag} when
electron passes the Si/Ge/Si heterointerfaces during tunneling
between QDs. To increase the spin relaxation time, one needs to
suppress the tunneling  by spatial separation of QDs.

The ESR study of the Si/Ge structure with QDs grown on Ge(001) shows
the only wide ESR signal that totally disappears after 650$^\circ$C
annealing. The narrow ESR signal from Si QDs grown on Ge(001) could
not be resolved plainly because of excessive signal broadening due
to a significant admixture of Ge into Si during the QD growth. The
GeSi intermixing is the result of Ge atoms segregation during  Si
deposition. By means this segregation the total energy of the system
decreases, since  Ge(001) surface has smaller surface energy than
Si(001) one~\cite{surfGe001}. In contrast, the Ge segregation during
Si growth on Ge(111) is strongly suppressed, because the difference
between Ge and Si surface energies for (111)-oriented samples is
much smaller~\cite{Si_Ge111}, making it possible to detect the ESR
signal from Si QDs grown on Ge(111).

\section{Conclusion}
In summary, the effective electron localization  in Si QDs was
confirmed by transport and ESR measurements. Temperature dependence
of the conductance is well described within a variable range hopping
model indicating a strong electron localization in the system. The
photoconductance behavior   is typical of the type-II QDs with one
type of localization carriers. The high Si content  in  QDs grown on
Ge(111) allows us to detect the narrow ESR signal from QD electrons.
It was shown that the isotropic ESR signal with $g$-factor
$g=2.0022$ and line width $\Delta H\approx 1.2$~Oe is related to the
electrons localized in QDs. An extrapolation of the empirical linear
dependence of the $g$-factor on the binding energy of localized
electrons in Si~\cite{Young} was used to predict the $g$-factor
value for electrons in  Si QDs. The experimental value is in a good
agreement with the $g$-factor estimated by taking into account the
strain effects and strong electron confinement in Si QD.

\begin{acknowledgments}
The authors acknowledge V.~A. Armbrister for Si QD growth,
A.~K.~Gutakovski for TEM study, A.~A.~Shklyaev for STM data and
V.~A.~Volodin for Raman measurements. The ESR study was funded by
Russian Scientific Foundation (grant No.~14-12-00931). The transport
measurements of Si QD structures were supported by RFBR (Grant
No.~13-02-00901).
\end{acknowledgments}

\end{document}